\documentclass{elsart}
\begin{document}
\begin{frontmatter}
\title{Study of continuum nuclear structure of \nuc{12}{C} via
        $(p,p'X)$ at intermediate energies}
\author{J. A. Templon\thanksref{caut}\thanksref{UGA}},
\author{B. A. Raue\thanksref{FIU}},
\author{L. C. Bland},
\author{K. Murphy\thanksref{SABBAGH}},
\author{D. S. Carman\thanksref{CMU}},
\author{G. M. Huber\thanksref{REGINA}},
\author{B. C. Markham},
\author{D. W. Miller},
\author{P. Schwandt}
\address{Indiana University Cyclotron  Facility,
         Bloomington, Indiana 47408, USA}
\author{D. J. Millener}
\address{Brookhaven National Laboratory, Upton, New York 11973, USA}
\thanks[caut]{Corresponding Author;
Dept.~of Physics and Astronomy \\
The University of Georgia, Athens, GA\ \ 30602\ \ USA \\
tel.~+1 706 542 2843 \ \ \ \ fax: +1 706 542 2492 \\
e-mail: {\tt templon@studbolt.physast.uga.edu}}
\thanks[UGA]{Present address: University of Georgia, Athens, GA 30602}
\thanks[FIU]{Present address: Florida International University,
    Miami, Florida 33199}
\thanks[SABBAGH]{Permanent address: Sabbagh Associates,
    Bloomington, Indiana 47408}
\thanks[CMU]{Present address: Carnegie-Mellon University,
    Pittsburgh, PA 15213}
\thanks[REGINA]{Permanent address: University of Regina,
    Saskatchewan, Canada}
\begin{abstract}
The inclusive $^{12}\mbox{C}(p,p')$ and exclusive $^{12}\mbox{C}(p,p'X)$
reactions have been studied with a beam energy of 156 MeV
and for $X = p \mbox{ and } \alpha$. The study focuses on the 
$(p,p'X)$ reaction mechanism and on the structure of $^{12}\mbox{C}$ 
just above the particle-emission threshold, $14 \leq E_x \leq 28$ MeV.
Cross sections were simultaneously measured for all three
reactions. The exclusive data were analyzed by making multiple-peak 
fits of the spectra and by Legendre-polynomial fits of the angular 
correlations.  Multiple-peak fits were also made of the inclusive
spectra.  The resultant cross sections were compared to
theoretical calculations. An analysis of the results
shows that this region of $E_x$ consists predominantly
of resonant excitations, in contradiction to
the findings of previous analyses.
\end{abstract}

\end{frontmatter}


Studies of structure in the nuclear continuum
suffer from difficulties in interpreting inelastic-scattering data.
The biggest problems are that the resonance peaks in the cross-section
spectra are often broad and poorly separated, and the amount of 
nonresonant ``background'' cross section underlying the resonance 
peaks is difficult to determine. Observation of the continuum 
decay products in coincidence with the inelastically-scattered 
probe can reduce these difficulties in two ways. First, since 
specific decay modes are accessible to only a subset of the
excited states, the density of states seen in a particular decay 
channel is lower than that seen in inclusive measurements.
Second, the angular distribution of the decay products contains
information about the reaction dynamics and can be used to judge 
the relative importance of non-resonant vs. resonant continuum 
excitation; for the latter, information about the angular 
momentum of the resonance excitations can be deduced.

These features have been beautifully realized in $(e,e'X)$ 
\cite{knop86} and $(\alpha, \alpha' X)$ \cite{zwar85} experiments,
but only a few $(p,p'X)$ experiments \cite{dera81,raue95} of
this type have been reported. Coincidence studies on $^{12}$C -
$(\alpha,\alpha'X)$ \cite{ried78}, $(p,p'\alpha)$ \cite{dera81},
$(e,e'p_0)$ \cite{cala84,cala94} and $(e,e'\alpha)$ \cite{dean95} -
have  concentrated on  searches for isoscalar giant resonance
strength. Since intermediate-energy proton 
inelastic scattering is an important source of our knowledge
of nuclear single-particle excitations, it is useful to 
examine the $(p,p'X)$ reaction to gain a better understanding 
of the continuum excited by the $(p,p')$ reaction.
This letter describes the first such intermediate-energy
experiment on a complex nucleus, $^{12}\mbox{C}$.
For states, or groups of states, up
to 20.6 MeV, these results may be compared with those from the
reactions $^{11}$B$(d,nX)$, where $X=p$ or $\alpha$ \cite{neus85}.


Data were accumulated for the reactions $^{12}\mbox{C} (p,p'X)$, 
where $X = p$ or $\alpha$, at the Indiana University Cyclotron 
Facility. Inclusive $(p,p')$ data were measured simultaneously.
A beam of 156 MeV protons with a DC current of 25--70 nA was 
focussed on a 2.0 mg/$\mbox{cm}^2$ natural-carbon target foil. 
Scattered protons were detected in the IUCF K600 spectrometer
\cite{berg86} at central scattering-angle settings of 14.3, 19, 
and 24 deg (momentum transfers $q_{\rm cm}$ of 0.71, 0.93, and 
1.16 $\mbox{fm}^{-1}$.) An energy resolution of 140 keV FWHM
was obtained, which is dominated by the energy spread of the 
incident beam. The proton scattering angle was reconstructed 
for each event to an accuracy of 0.1$^\circ$. Coincident 
low-energy charged particles were detected in an array of 
eight silicon detector telescopes arranged to provide maximal 
coverage of the in-plane angular range. The residual-nucleus 
mass resolution was better than 250 keV FWHM. For $^{12}\mbox{C}$ 
excitation energies $14 \leq E_x \leq 28$ MeV, $\alpha$-particle 
emission to both the ground and first-excited states of
$^{8}\mbox{Be}$ were observed, and proton emission was observed
to the ground state and several excited states of $^{11}\mbox{B}$.
Only the two most prominent decay channels --- $\alpha_1$
(to the $J^\pi = 2^+$ first excited state of $^{8}\mbox{Be}$) 
and $p_0$ (to the $J^\pi = 3/2^-$ $^{11}\mbox{B}$ ground state) 
---  are discussed below. 


Fig.~\ref{fig:channels} shows representative spectra for the 
inclusive and exclusive reactions at $q = 0.7\mbox{ fm}^{-1}$.
Two observations are apparent from these spectra. The first is 
that the separation of the continuum spectrum into specific 
decay channels, by requiring the detection of specific coincident 
decay products, gives the expected reduction of level density.
The $p_0$ and $\alpha_1$ gated spectra are both less complex than 
the inclusive spectrum and have few common features.
The second observation is that the coincidence spectra do not
display the large nonresonant backgrounds which are normally
associated with inclusive spectra.
There is little evidence in the coincidence data for
{\em any\/} background for $E_x \leq 24$ MeV.
A similar observation has been made in the study of the
$^{12}\mbox{C}(e,e'p)$ reaction \cite{wojt93}.
These observations led us to two conjectures:
1) the $^{12}\mbox{C}$ continuum at low excitation energy is
less complex than commonly believed and can be analyzed
in terms of individual excitations, and
2) the common assumption of large nonresonant contributions
to the $^{12}\mbox{C}(p,p')$ spectrum in the low-energy (15--25 MeV)
continuum is incorrect.

The level-density reduction is best demonstrated in the 
$(p,p'\alpha_1)$ spectrum. One can clearly see four peaks 
below 24 MeV: the well-known states \cite{ajze90} at
$E_x$ = 15.4 MeV ($J^\pi = 2^+, T=0$), 16.1 MeV 
($J^\pi = 2^+, T=1$), 18.3 MeV ($J^\pi = 2^-, T=0$),
and 21.6 MeV ($J^\pi = 2^+, T=0$).
The 16.1 MeV $T=1$ state is known to be isospin-mixed
and to decay mainly into the $\alpha_1$ channel \cite{ajze90}.
The lack of background strength is illustrated by the $(p,p'p_0)$
spectrum
in the inset of Fig.~\ref{fig:channels}.
The region shown (17--21 MeV) has been well studied
\cite{morr79,neus83,buen77,comf82,jone83,hick84}
and is dominated at this momentum transfer by states at 
$E_x$=18.3 MeV (the same mentioned above for the $\alpha_1$
channel), 19.4 MeV ($J^\pi=2^-$,$T=1$) and 20.6 MeV 
(unknown spin and parity).
The curve shows the result of a least-squares fit
which assumed the presence of only these three peaks,
without inclusion of any background. The inclusive spectrum, 
on the other hand, suggests the presence of significant 
nonresonant backgrounds. Large backgrounds have been included 
in earlier analyses of $(p,p')$ spectra in this region of 
excitation \cite{buen77,comf82,jone83} 
(see especially \cite{jone84}).

The $(p,p'\alpha_1)$ spectrum suggests a possible resolution 
of this background disparity. The long lorentzian tails of 
the broad states at 15.4 and 21.6 MeV are clearly visible, 
and contribute significantly to the yield under the peak at 
18.3 MeV. Failure to properly represent these broad resonances
when fitting the inclusive spectrum would lead to a false 
identification of a large nonresonant background. This 
conclusion might even be drawn if these resonances were
included but had incorrect widths. The studies 
\cite{comf82,jone83,comf80}, in which gaussian shapes have 
been used, provide examples of this effect and result in 
underestimation of the cross section.

One of the motivations for this experiment was to measure 
the angular correlation functions (ACFs) and to determine 
the extent of their sensitivity to the angular momentum of 
the resonance excitations. The ACF is simply the cross section
${\rm d}^5\sigma/{\rm d}\Omega_{p'} {\rm d}\Omega_X {\rm dE}_X$ 
plotted as a function of $\theta_{X}^{\rm c.m.}$ (at fixed 
$\theta_{p'}$), defined in the center-of-mass frame of the 
recoiling $A=12$ system. This is the natural frame for a 
multipole decomposition of the ACF. The angle 
$\theta_{X}^{\rm c.m.} = 0$ corresponds to the $(p,p')$
momentum-transfer direction $\hat{q}$.
Under the assumptions that the reaction proceeds by sequential
resonance excitation and decay, and that the ACFs are
independent of the azimuthal angle $\phi_{X}^{\rm c.m.}$, the 
ACFs can be described by a Legendre-polynomial ($P_\ell$) series 
\cite{temp93}. In regions dominated by a single resonance, 
this model predicts that only even-$\ell$ terms with 
$\ell_{\rm max} \leq 2J$, where $J$ is the total angular 
momentum of the resonance, are needed in the fit.

Fig.~\ref{fig:legfits} displays several ACFs, for the
$p_0$ and $\alpha_1$ channels, for $E_x(^{12}\mbox{C})$
values centered on three of the observed resonances.
The ACFs are different for the three regions, indicating a
sensitivity to nuclear structure. A single resonance appears 
to dominate the region around 21.6 MeV in the $(p,p'\alpha_1)$ 
channel. Several experiments (\cite{ried78,dean95} and 
references therein) assign $J^\pi = 2^+,\ T=0$ to this level.
The Legendre polynomial fit shown in the figure has large reduced
coefficients ($b_\ell$) for $P_2$ and $P_4$, and those for 
$P_1$ and $P_3$ are an order of magnitude smaller, consistent 
with the $2^+$ assignment. The other prominent peak in the
$(p,p'\alpha_1)$ spectrum at E$_x$=24.4 MeV has an ACF similar 
in shape to the 21.6 MeV resonance, suggesting that it too is 
a $J^\pi = 2^+,\ T=0$ level.

The experiments of \cite{zwar85,suko87} have demonstrated that
significant quasifree knockout strength leads to large 
fore-aft asymmetries in the ACFs. No such asymmetry is seen 
in the two ACFs at lower $E_x$. This further supports the 
conjecture that little nonresonant background exists in this region.

The channel cross section at a given $E_x$ is given by the 
integral over $\d \Omega_X$ of the ACF. If the assumption 
of $\phi_{X}^{\rm c.m.}$-independence for the ACFs holds,
this cross section is given by $4\pi a_0$, where $a_0$ is 
the coefficient of $P_0(\theta)$ in the Legendre-polynomial 
fit to the data.
These cross sections in turn indicate the channel composition of
the continuum seen by inclusive $^{12}\mbox{C}(p,p')$.
The deduced angle-integrated cross sections
for $\alpha_1$ and for $p_0$  account for respectively 24\% 
and 36\% of the inclusive cross section for 
$17 \leq E_x \leq 24$ MeV. 
The corresponding $n_0$ contribution over this region cannot be
deduced from the data;
only $s$-wave emission will be important close to the threshold
at 18.72 MeV, with $d$-wave
emission rising to 50\% of $p_0$ at about 22.7 MeV.
However, estimates can be made for specific cases.
For the peak at 19.4 MeV, $p_0$ and $\alpha_1$ emission account
for 48\% and 20\% of the inclusive cross section at low $q$,
where $2^{-}$ strength is known to dominate.
The shell-model $n_0$ width for this $2^{-}$ state
should be  enhanced by
the isospin mixing observed in pion scattering \cite{morr79}.
A 5\% isospin mixing by intensity, chosen to fit the ratio
of $(e,e')$ cross sections \cite{hick84}, of the second $2^-$
shell-model states with $T=0$ and $T=1$ raises the $n_0$ contribution
from 16\% to 40\%, which means that all the inclusive cross section
is accounted for.
The measured width for this state (see Table \ref{tab:compare})
is well reproduced by the calculated nucleon decay width of
468 keV ($\Gamma_{p_0}(d) = 86$ keV, $\Gamma_{p_0}(s) = 168$ keV 
and $\Gamma_{n_0}(s) = 214$ keV) and the measured $\alpha_1$
width ($\sim 100$ keV.)

The conjectures about the $(p,p')$ continuum, reinforced by this
quantitative analysis of the $(p,p'X)$ coincidence data, necessitate
a reanalysis of the inclusive data. This analysis was performed 
on the present $(p,p')$ data sorted into twelve spectra, each
corresponding to a one-degree bin of scattering angle $\theta_{p'}$.
The twelve spectra were fit simultaneously using identical peak
centroids and widths. The background contribution was represented 
in the fit by a lorentzian function whose centroid, width, and 
area varied freely at each angle. It is likely that this background 
represents the low-energy tails of higher-excitation resonances.  
Broader resonance structures are clearly evident in the 
$(p,p'\alpha_1)$ spectrum in Fig.\ \ref{fig:channels}. A more 
quantative treatment of this excitation region would require 
response functions from a continuum shell model calculation since 
the broad structure near 25 MeV cannot be represented by a single 
lorentzian lineshape. Fig.\ \ref{fig-incl-peakfit} shows the 
results of this analysis for one of the twelve spectra.
Comparable representations were obtained for all spectra using
a fit that includes thirteen resonances and
one background function.
The peak parameters for the most prominent of these resonances are
tabulated in Table\ \ref{tab:compare}.
Also listed are the corresponding data of \cite{buen77} for
comparison.
The errors on the peak positions and widths are small since
a weighted mean has been made of the results for each of the
twelve spectra for most cases.
The correlations between the parameters in the fit are
only partially
included in these errors.

A comparison of the current inclusive analysis with
previous data provides information about the effect
of background overestimation on the extracted cross sections.
Only the pioneering work of Buenerd {\em et al.\/}
\cite{buen77} reports cross sections over the 
range 14--25 MeV, although the data of Comfort {\em et al.\/}
\cite{comf82,comf81} extend to the lower edge of the
current experiment's energy acceptance.

Fig.\ \ref{fig:bigfig} displays cross section results for several
states along with results from the previous experiments.
The data of \cite{comf82} for the states at 18.3, 19.4, and
19.7 MeV are from a 200 MeV experiment; Comfort has shown
previously \cite{comf81} that $(p,p')$ cross sections over the 
range 120-200 MeV have a weak beam-energy dependence.
The three datasets are in good agreement for the sharp
states at 15.1 ($J^\pi = 1^+,\ T = 1$) and 16.1 
($J^\pi = 2^+,\ T = 1$) MeV (not shown), and with half
the cross sections for the corresponding states in
$(p,n)$ and $(n,p)$ reactions (\cite{ande96} and references
therein). For both states, 
the current data are identical within errors to that of 
\cite{comf81}, while those of \cite{buen77} agree in shape 
but are lower by a factor 1.6. For the remaining states, 
the data of \cite{buen77} agree well in shape with the 
current results, but usually not in magnitude.
Since Comfort
{\it et al.} \cite{comf82} made no attempt to decompose
the strong peak at 19.7 MeV into $2^-$ and $4^-$ components,
comparisons should be made for the summed strength in the
19.4-MeV and 19.7-MeV panels in Fig.\ \ref{fig:bigfig}.
The figure illustrates nicely how 
resonance cross-section results depend strongly on the 
assumptions of the peak-fitting procedure, and how previous 
experiments have tended to underestimate the cross sections.
With two exceptions, our cross sections are everywhere 
larger or equal to those of the other two experiments.

A comparison with the results of nucleon charge-exchange reactions
corroborates the conclusions reached above.
Simple isospin-symmetry arguments predict that the $(p,p')$
cross sections for a given peak should be at least half that of
the corresponding peak in $(p,n)$ or $(n,p)$ reactions.
For the region around 19.5 MeV
at the lowest  $q$ measured, near the
maximum for the $2^-$ T=1 state which dominates, the $(p,n)$
cross section of \cite{ande96}, also  extracted using
Lorentzian lineshapes, is equal to  twice the summed cross section
for the 19.4-MeV and 19.7-MeV peaks in Fig.\ \ref{fig:bigfig}.
The cross sections of Comfort at 200 MeV \cite{comf82} and Jones
\cite{jone83} at higher energies fall well short of expectations 
based on  the charge-exchange data, the more so of other recent
data \cite{yang93} and several older data sets (see references in
\cite{ande96}). Given the similar backgrounds subtracted for
the 19.4-MeV and 18.3-MeV peaks \cite{jone84}, the discrepancies
seen in Fig.\ \ref{fig:bigfig} for the 18.3-MeV peak are not
surprising.

 DWIA calculations were compared with the new inclusive cross 
sections in order to associate the resonances with levels 
from shell-model calculations \cite{hick84,brad91,ande96}
using interactions from \cite{cohe65,mill75}.
Calculations were performed using the code DW86 \cite{scha70},
the NN effective interaction of \cite{fran85}, the shell-model
one-body density-matrix elements with harmonic oscillator
wave functions ($b_0 = 1.669$ fm or $b_{rel} = 1.743$ fm
\cite{brad91,ande96}) and the $^{12}\mbox{C}$ optical potential 
of \cite{comp74}. The results of the calculations are displayed 
in Fig.\ \ref{fig:bigfig}; any normalization factors by which 
the curves are scaled are shown in parentheses.
The normalization factors are also listed in Table \ref{tab:compare}.
When a specific transition density is identified
with a peak, the subscript indicates which of the several states
for each $(J^\pi , T)$ was used. The agreement in most cases is 
good, and the normalization factors have reasonable sizes in the sense
that quenching of the shell-model transition densities is expected
for all the cases listed in Table \ref{tab:compare} 
\cite{brad91,ande96}.

 Two curves for $J^\pi = 4^-$ excitation are shown with the data for
the state at 19.7 MeV. Individually, the normalization factors for 
these curves are twice as large as those characteristic of the 
other states but the normalization factor of 0.55 for the sum
of the T=0 and T=1 $4^-$ states is consistent with that required
for the T=1 state in the $(p,n)$ reaction \cite{ande96}.
A strongly isospin-mixed $J^\pi = 4^-$ doublet, with peaks at 
19.25 and 19.65 MeV, was observed in a $\pi^+/\pi^-$ 
inelastic-scattering experiment \cite{morr79}. No peak at 19.25 
MeV was observed in the current experiment. Its absence suggests
that the isospin mixing results in all of the $(p,p')$ strength 
going to the 19.7 MeV state. However, DWIA calculations for
a more proton-like lower state do not support this hypothesis. 
The fact that the $4^-_2$ and $4^-_3$ T=0 shell-model states are 
closely spaced and share the $L=3$ $S=1$ excitation strength 
may complicate the isospin mixing calculation.

The level at 20.6 MeV has been one of the most difficult states
to explain in \nuc{12}{C} (see \cite{neus83,comf82,hick84,ande96} and
references therein.) States with $J^\pi = 3^+$ and $3^-$, both with $T=1$,
are certainly present \cite{ajze90} but the $(p,n)$ results of 
\cite{ande96} suggest that T=1 states account for only 25\% of the 
cross section shown in Fig. \ref{fig:bigfig}. Neither of these
states can account for the large $^{11}$B$(d,n)$ cross section
for the 20.6-MeV peak \cite{neus83}. However, the 
$J^\pi = 3^-_4, T=0$ state included in Fig. \ref{fig:bigfig}
is predicted to lie in the energy region and has a large 
ground-state spectroscopic factor of 0.56 (mainly $d_{3/2}$).
None of these calculations reproduce the data alone, although the sum
of the three has at least the correct order of magnitude.
Recent $(\pol{d},\pol{d}')$ experiments \cite{john95} indicate an
isoscalar $J^\pi = 1^+$ resonance at 20.5 MeV, but none of the
unassigned $1^+,T=0$ wavefunctions of \cite{cohe65}
have significant $(p,p')$ strength.
The nature of the states in this region remains unclear.

Evidence was presented above which was consistent with a $J^\pi = 2^+ \ 
T=0$ assignment for the 21.6 MeV state. For this state,
as well as for the 15.4-MeV $2^+$ T=0
state, neither the large inclusive cross section nor the large postive 
analysing power (not shown) over the $q$ range measured can be 
reproduced by a $0\hbar\omega$ \cite{cohe65}
shell-model wavefunction.
Essentially all the $0\hbar\omega$ $E2$ strength is contained
in one state, which corresponds to the 4.44-MeV level.
A modest fraction of the $2\hbar\omega$
giant-quadrupole strength built on the $0\hbar\omega$ ground state
is required to explain the cross sections and analysing powers -
roughly 16\% in the case of the of the  21.6 MeV state.


In summary, a $^{12}\mbox{C}(p,p'X)$ experiment has been performed
at a beam energy of 156 MeV.
Analysis of the coincidence data indicates
that the nonresonant component of the low-energy
($14 \leq E_x < 24$ MeV) continuum
is much smaller than commonly accepted.
The measured angular correlation functions follow the pattern
expected for a resonance excitation-decay process, which
also points to relatively small nonresonant contributions.
The ACFs display sensitivity to nuclear structure.

The inclusive $(p,p')$ data, measured simultaneously in this experiment, 
were analyzed using more physically motivated peak shapes than used in
earlier analyses.  This reanalysis produces cross sections that are,
in general, much larger than deduced from previous $^{12}\mbox{C}(p,p')$
experiments.  Theoretical calculations agree quite well with the
present data and require renormalization factors which are
qualitatively in line with the quenching expected and to those
needed for the corresponding T=1 states excited in recent
charge-exchange reactions. At the low momentum transfers
probed in the present experiment, we have deduced that the low-energy
($21 < E_x < 24$ MeV) continuum of $^{12}$C is dominated by
1$^-$, T=1 resonances which decay primarily by single-nucleon 
emission, and 2$^+$, T=0 resonances
which decay primarily by alpha emission.  There is little, if any, 
non-resonant background in the spectrum.  This is a qualitatively
different view of the continuum than has been presented in earlier works.  
Instead of viewing the continuum as giant resonances sitting atop
a sizable non-resonant background, the present analysis suggests
that {\it most} of the continuum cross section is due to resonance
excitation, with clear peak structures melting into the familiar
smooth continuum as the width of the overlapping resonances increases
with increasing excitation energy.

\begin{ack}

We thank
Dr.\ J.\ Comfort for providing a tabulation of his data.
This work was financially supported by the National Science Foundation
and  by the US Department of Energy under Contract No. 
DE-AC02-76CH00016.
One of the authors (J.A.T.) was funded by a National
Science Foundation Graduate Fellowship during a portion of this project.

\end{ack}


\newpage

\begin{figure}
\caption{Representative \nuc{12}{C}$(p,p')$ and \nuc{12}{C}$(p,p'X)$
spectra (counts {\em vs.\/} excitation energy $E_x$) at
proton scattering angle $\theta_{p'} = 14.3^\circ$.
The coincidence data shown are for a correlation angle
$\theta_X^{\rm c.m.} = 0^\circ$ (along \protect{$\pol{q}$}).}
\label{fig:channels}
\end{figure}

\begin{figure}
\caption{Representative ACFs for the $p_0$ and $\alpha_1$ decay
channels. The solid curves are Legendre-polynomial fits.
In all cases $\ell_{\rm max} \leq 4$.
Coefficients of polynomials with larger $\ell$ values are not
sufficiently constrained by the data's
$25^\circ$ spacing in $\theta_{X}^{\rm c.m.}$.
The first two sets of ACFs correspond to the $J^\pi = 2^-, T=0$ (18.3 MeV),
$J^\pi = 2^+, T=0$ (21.8 MeV) resonances.
Reduced coefficients for the Legendre-polynomial fit are shown
for the 21.8 MeV $(p,p'\alpha)$ ACF, with the estimated
errors in parentheses.
The third set is centered about a resonance at 24.4 MeV which is only
seen in the $(p,p'\alpha_1)$ channel.}
\label{fig:legfits}
\end{figure}

\begin{figure}
\caption{Multiple-peak fit of the inclusive spectrum for 
$\theta_{p'} = 14.3^\circ$ (lab).
The area under
the curve labelled B is the background contribution.}
\label{fig-incl-peakfit}
\end{figure}

\begin{figure}
\caption[a]{Differential cross sections for selected transitions.
The curves are DWIA calculations discussed in the text.
For  the  15.1-MeV $1^+$ state, the curve labelled FIT uses
a $p$-shell transition density optimized to fit the measured
$(e,e')$ form factor \cite{brad91}.
The error bars are smaller than the symbols for some data points.
Filled circles: present experiment; diamonds: Comfort 
{\em et al.} \cite{comf82,comf81}; squares: 
Buenerd {\em et al.} \cite{buen77}.}
\label{fig:bigfig}
\end{figure}

\newpage

\begin{table}
\caption[Big Table]{Comparison of current and previous continuum
  $^{12}\mbox{C}(p,p')$ results (see text.) The subscripts on the 
  spin $J$ indicate the specific wavefunction of \cite{cohe65,mill75}
  which gave the best fit to the data in the DWIA calculation.
  Quantum numbers without a subscripted $J$ are suggested by
  systematics or determined in other experiments \cite{ajze90}.
  ``DWIA Scaling'' refers to the best-fit multiplicative
  factor  applied to DWIA calculations.}
\label{tab:compare}
\begin{tabular}{cccccc} \hline
\multicolumn{4}{c}{Current Experiment} & 
	\multicolumn{2}{c}{Ref.~\protect\cite{buen77}} \\ \hline
$E_x$  &  $\Gamma$  & $J^\pi \ \ (T)$ & {DWIA} &  $E_x$  &  $\Gamma$  \\
(MeV $\pm$ keV) & (keV) & \null & \multicolumn{1}{c}{Scaling} &
	(MeV $\pm$ keV) & (keV) \\ \hline
$15.38 \pm 30$  & $2800 \pm 170$ & $2^+\ \ (0)$   & ---  & $15.3  \pm 200$ & $2000 \pm 200$ \\
$16.62 \pm 10$  & $280  \pm 30$  & $2^-\ \ (1)$   & ---  & ---             & ---            \\
$18.292 \pm 4$  & $486  \pm 10$  & $2^-_2\ \ (0)$ & 0.80 & $18.35 \pm 50$  & $400  \pm 100$ \\
$19.394 \pm 10$ & $520  \pm 30$  & $2^-_2\ \ (1)$ & 0.38 & $19.4  \pm 50$  & $530  \pm 100$ \\
$19.671 \pm 6$  & $490  \pm 20$  & $4^-\ \ (0+1)$ & 0.58 & $19.6  \pm 50$  & $500  \pm 100$ \\
$20.584 \pm 5$  & $440  \pm 11$  & ---            & ---  & $20.6  \pm 80$  & $450  \pm 150$ \\
$21.61  \pm 20$ & $1450 \pm 90$  & $2^+\ \ (0)$   & ---  & $21.3  \pm 250$ & $950  \pm 300$ \\
$21.99  \pm 20$ & $550  \pm 90$  & $1^-_3\ \ (1)$ & 0.78 & $21.95 \pm 150$ & $800  \pm 100$ \\
$22.72  \pm 30$ & $1200 \pm 130$ & $1^-_4\ \ (1)$ & 0.60 & $22.6  \pm 150$ & $900  \pm 100$ \\
$23.57  \pm 20$ & $238  \pm 34$  & $1^-\ \ (1)$   &      & $23.50$         & $230$          \\
$24.04  \pm 18$ & $659  \pm 48$  & $1^-\ \ (1)$   & ---  & $23.92$         & $400$          \\
$24.38  \pm 10$ & $671  \pm 49$  & $2^+\ \ (0)$   & ---  & ---             & ---            \\ \hline
\end{tabular}
\end{table}
\vfill

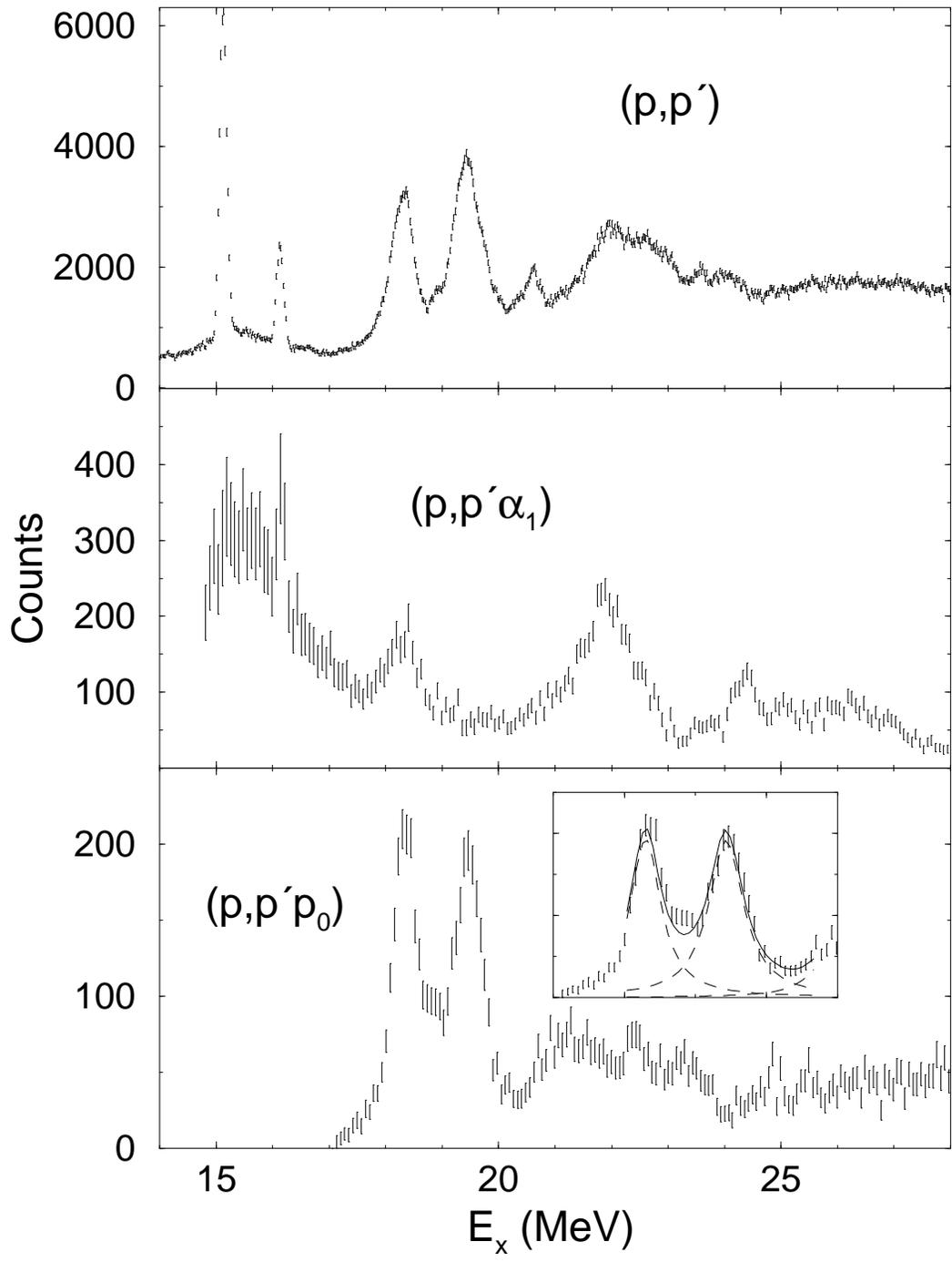
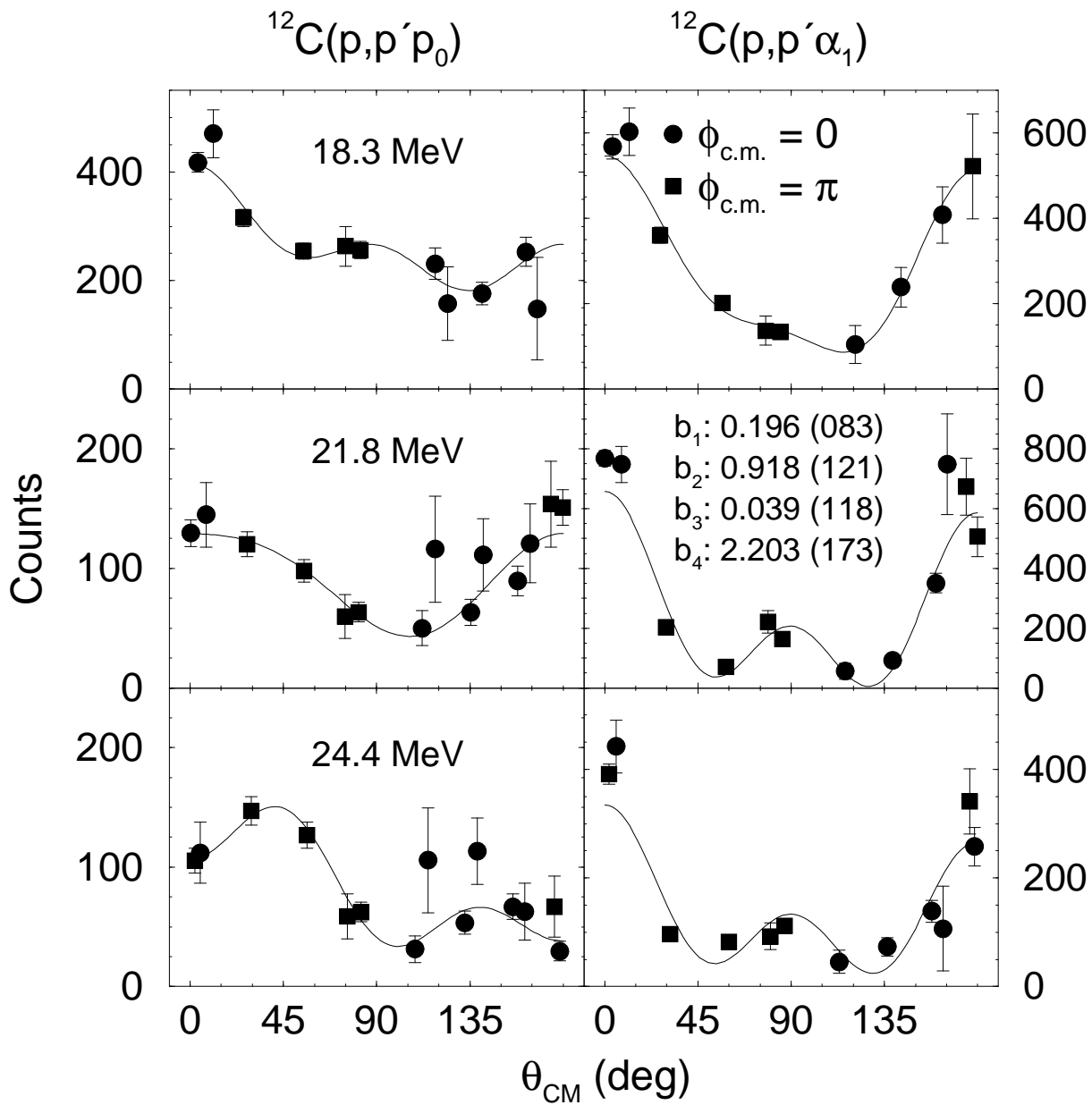
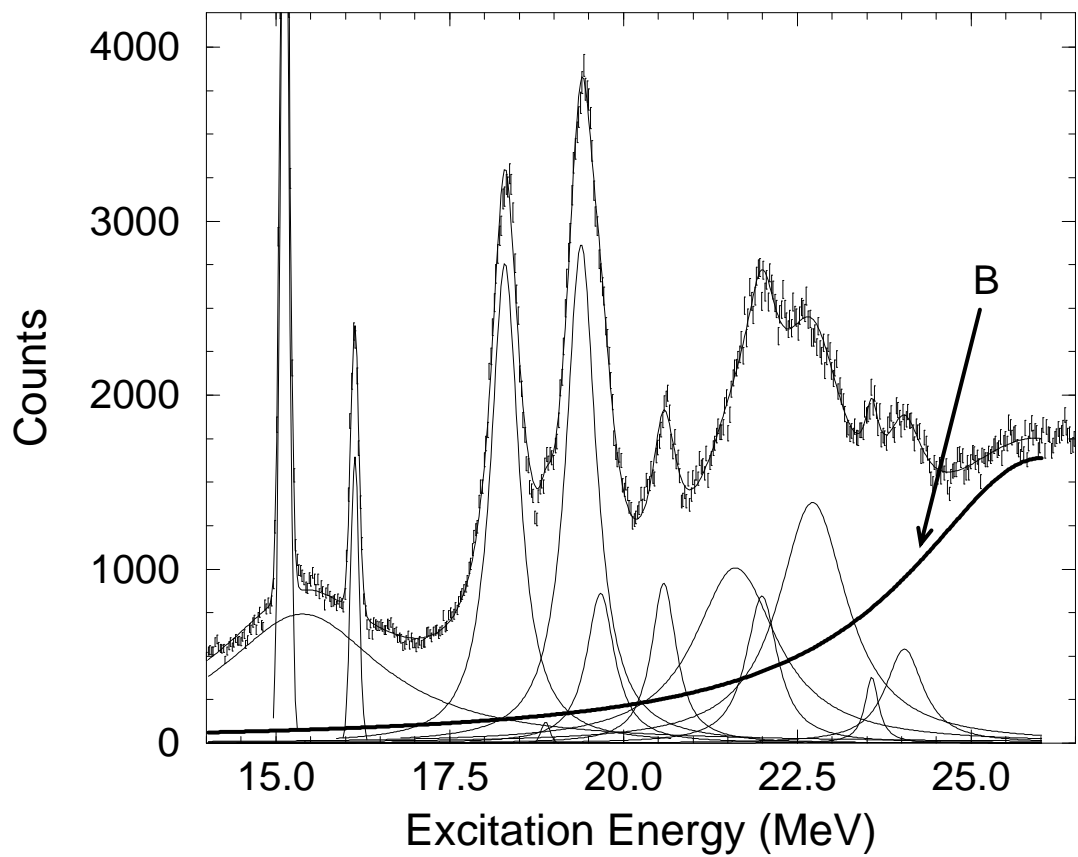
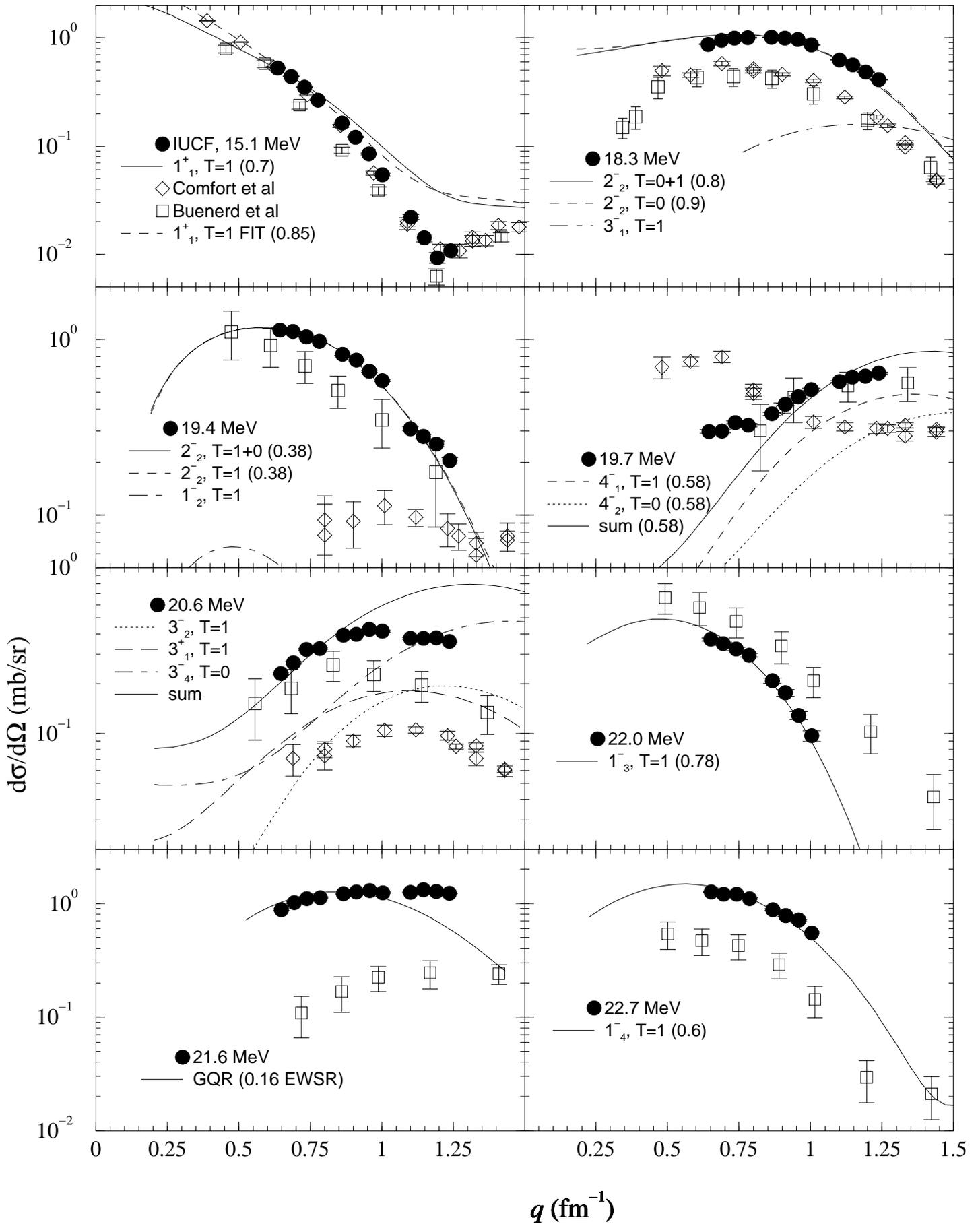
\end{document}